\documentclass[conference]{IEEEtran}
\IEEEoverridecommandlockouts
\usepackage{cite}
\usepackage{amsmath,amssymb,amsfonts}
\usepackage{algorithmic}
\usepackage{graphicx}
\usepackage{textcomp}
\usepackage{xcolor}
\usepackage{dblfloatfix} 
\def\BibTeX{{\rm B\kern-.05em{\sc i\kern-.025em b}\kern-.08em
    T\kern-.1667em\lower.7ex\hbox{E}\kern-.125emX}}

\begin{document}

\title{Spatio-Temporal Coverage Enhancement in Drive-By Sensing Through Utility-Aware Mobile Agent Selection\\
}

\author{\IEEEauthorblockN{Navid Hashemi Tonekaboni\IEEEauthorrefmark{1},
Lakshmish Ramaswamy\IEEEauthorrefmark{2}, Deepak Mishra\IEEEauthorrefmark{4}, Sorush Omidvar\IEEEauthorrefmark{3}}\\
\IEEEauthorblockA{\IEEEauthorrefmark{1}\IEEEauthorrefmark{2}Department of Computer Science, \IEEEauthorrefmark{4}Department of Geography, \IEEEauthorrefmark{3}College of Engineering\\
\IEEEauthorrefmark{1}Emory University, Atlanta, GA 30322, USA\\
\IEEEauthorrefmark{2}\IEEEauthorrefmark{4}\IEEEauthorrefmark{3}University of Georgia, Athens, GA 30602, USA\\
\IEEEauthorrefmark{1}navid.ht@emory.edu,
\IEEEauthorrefmark{2}laks@cs.uga.edu,
\IEEEauthorrefmark{4}dmishra@uga.edu,
\IEEEauthorrefmark{3}omidvar@uga.edu}}

\maketitle

\begin{abstract}
In recent years, the drive-by sensing paradigm has become increasingly popular for cost-effective monitoring of urban areas. Drive-by sensing is a form of crowdsensing wherein sensor-equipped vehicles (aka, mobile agents) are the primary data gathering agents. Enhancing the efficacy of drive-by sensing poses many challenges, an important one of which is to select non-dedicated mobile agents on which a limited number of sensors are to be mounted. This problem, which we refer to as the mobile-agent selection problem, has a significant impact on the spatio-temporal coverage of the drive-by sensing platforms and the resultant datasets. The challenge here is to achieve maximum spatiotemporal coverage while taking the relative importance levels of geographical areas into account. In this paper, we address this problem in the context of the SCOUTS project, the goal of which is to map and analyze the urban heat island phenomenon accurately.

Our work makes several major technical contributions. First, we delineate a model for representing the mobile agents selection problem. This model takes into account the trajectories of the vehicles (public transportation buses in our case) and the relative importance of the urban regions, and formulates it as an optimization problem. Second, we provide two algorithms that are based upon the utility (coverage) values of mobile agents, namely, a hotspot-based algorithm that limits the search space to important sub-regions and a utility-aware genetic algorithm that enables the latter algorithm to make unbiased selections. Third, we design a highly efficient coverage redundancy minimization algorithm that, at each step, chooses the mobile agent, which provides maximal improvement to the spatio-temporal coverage. This paper reports a series of experiments on a real-world dataset from Athens, GA, USA, to demonstrate the effectiveness of the proposed approaches.\end{abstract}

 
\section{INTRODUCTION}

The massive proliferation of mobile sensor devices is changing the landscape of environmental monitoring by augmenting conventional data sources such as satellites and weather stations with the crowdsensing paradigm. In particular, crowdsensing is very beneficial in urban areas where higher population densities not only provide larger pools of potential contributors but also enhances the impact of crowdsensing on the local population. 

Although people are considered to be the main participants in crowdsensing, a new category of this paradigm, namely \textit{drive-by sensing}, has recently emerged. In the drive-by sensing paradigm, the primary sensing agents are vehicle-borne sensors \cite{lee2010survey}. Drive-by sensing has numerous applications in urban and environmental monitoring, especially where the properties that are being monitored exhibit strong spatio-temporal associations. Google street view \cite{anguelov2010google} is a famous example of the drive-by sensing paradigm, wherein vehicles are employed to collect street-level imagery on a global scale. Drive-by sensing paradigm has also been employed to monitor the road conditions at both the surface and sub-surface levels \cite{wang2014framework}.

A recent research direction has been to employ non-dedicated vehicles (vehicles whose primary functionality is not sensing/data gathering) for drive-by sensing. Here, sensing occurs \textit{opportunistically} during the regular operation of the vehicles. For example, in the city scanner project, sensors were mounted on municipal garbage trucks to collect a multitude of environmental parameters of the city without interfering with the routes or operations of the truck fleet \cite{anjomshoaa2018city}. 
 
Drive-by sensing through public transportation vehicles (e.g., city buses) is attractive because of the several advantages it offers. First, these vehicles move around the cities frequently throughout the day, providing a cost-effective means to monitor large swathes of cities. Second, because these vehicles ply on pre-defined routes and follow pre-defined schedules, it is possible to estimate their locations at a given time of the day. This permits systematically planned data gathering. Since these vehicles return to specific locations at the end of their shifts, scheduling the mounting and maintenance of sensors becomes less cumbersome.

While drive-by sensing through non-dedicated vehicles is becoming popular, making it effective, efficient and practical poses significant difficulties such as the lack of control on the routes and schedules of these vehicles, uneven spatio-temporal sensing coverage and the high costs and human efforts involved in installing and maintaining the sensors on vehicles. One of the major research challenges is to select a subset of public transportation vehicles (i.e., buses) to mount the sensor devices. This challenge acquires importance because budgetary constraints and human efforts required in installing and maintaining sensors often limit the number of sensors that can be deployed (i.e., it is impractical to deploy sensors on all buses of the city). For instance, a single sensor setup to monitor urban temperatures costs more than a hundred dollars, and installing and configuring a setup on a bus requires a few hours of work from a human expert.

In this context, it is imperative to design a cost-effective strategy to select buses for installing a limited number of sensors so as to maximize the benefits of sensing in terms of spatio-temporal granularity and coverage of the sensed data. In this paper, we refer to this as the \textit{mobile agent selection problem}. Furthermore, based upon the urban phenomena being monitored, certain parts of the city may have a higher importance in the sense that they may need to be sensed at higher spatio-temporal resolutions. For example, these may be densely populated regions or regions with significant variations in the environmental parameters. Thus, the mobile agent selection problem (city buses in the context of this study) has to take into account the relative importance levels of the sub-regions of an urban area. 

In this paper, we focus on this problem in the context of the SCOUTS project \cite{tonekaboni2018scouts}, the goal of which is to generate hyperlocal heatmaps of urban regions with high spatio-temporal granularity. In addressing this problem, we make several novel technical contributions, which can be summarized as follows:

\begin{itemize}
\item We provide a novel mathematical formulation for the mobile agent selection problem. This model is unique in the sense that it takes into account the trajectories of vehicles as well as the relative importance levels of various sub-regions of an urban area. We formulate this as a constrained optimization problem with an objective function that encapsulates spatio-temporal sensing granularity requirements. 
\item We propose two algorithms, namely the hotspot-based algorithm and the utility-aware genetic algorithm. The hotspot-based algorithm shrinks the spatial grid of the whole city to a grid of hotspot cells, which have higher relative importance compared to other grid cells. The genetic algorithm is founded on top of the hotspot-based algorithm with a configuration to make unbiased selections. Considering that reducing the whole grid to only the grid of hotspot cells might result in an unfair agent selection, the latter algorithm provides the sensing agents, which have not covered any hotspot area, with a chance to be selected.

\item Third, we design a highly efficient redundancy minimization algorithm. At each step, this algorithm chooses the bus that provides maximal improvement to the spatio-temporal coverage of the current selection. This is done by minimizing the redundancies caused by overlapping trajectories. This algorithm not only outperforms all of the above algorithms in terms of the spatio-temporal sensing coverage but also runs orders of magnitude faster than an exhaustive search approach. \end{itemize}

We evaluate the performance of all the proposed algorithms on a real-world bus trajectory dataset from the public transit system from Athens, Georgia, USA (ACC public transit). Our experiments show that our proposed algorithms significantly enhance the spatio-temporal coverage of a limited number of sensing agents.

\section{BACKGROUND AND MOTIVATION}
\subsection{SCOUTS Project}
The SCOUTS project at the University of Georgia aims to generate hyperlocal heat exposure maps of different urban communities \cite{tonekaboni2018scouts}. For this purpose, we augment traditional data sources such as satellites and weather stations with human-borne crowdsensing and drive-by sensing. Public transportation buses were deemed to the most suitable agents for drive-by sensing because of two reasons. On the one hand, the routes of these buses are close to the daily commute of city-dwellers. On the other hand, the constant movement of these vehicles throughout the cities provide reasonable spatio-temporal coverage of urban regions. 

\begin{figure}[htbp]
    \centering
    \includegraphics[width=0.6\columnwidth]{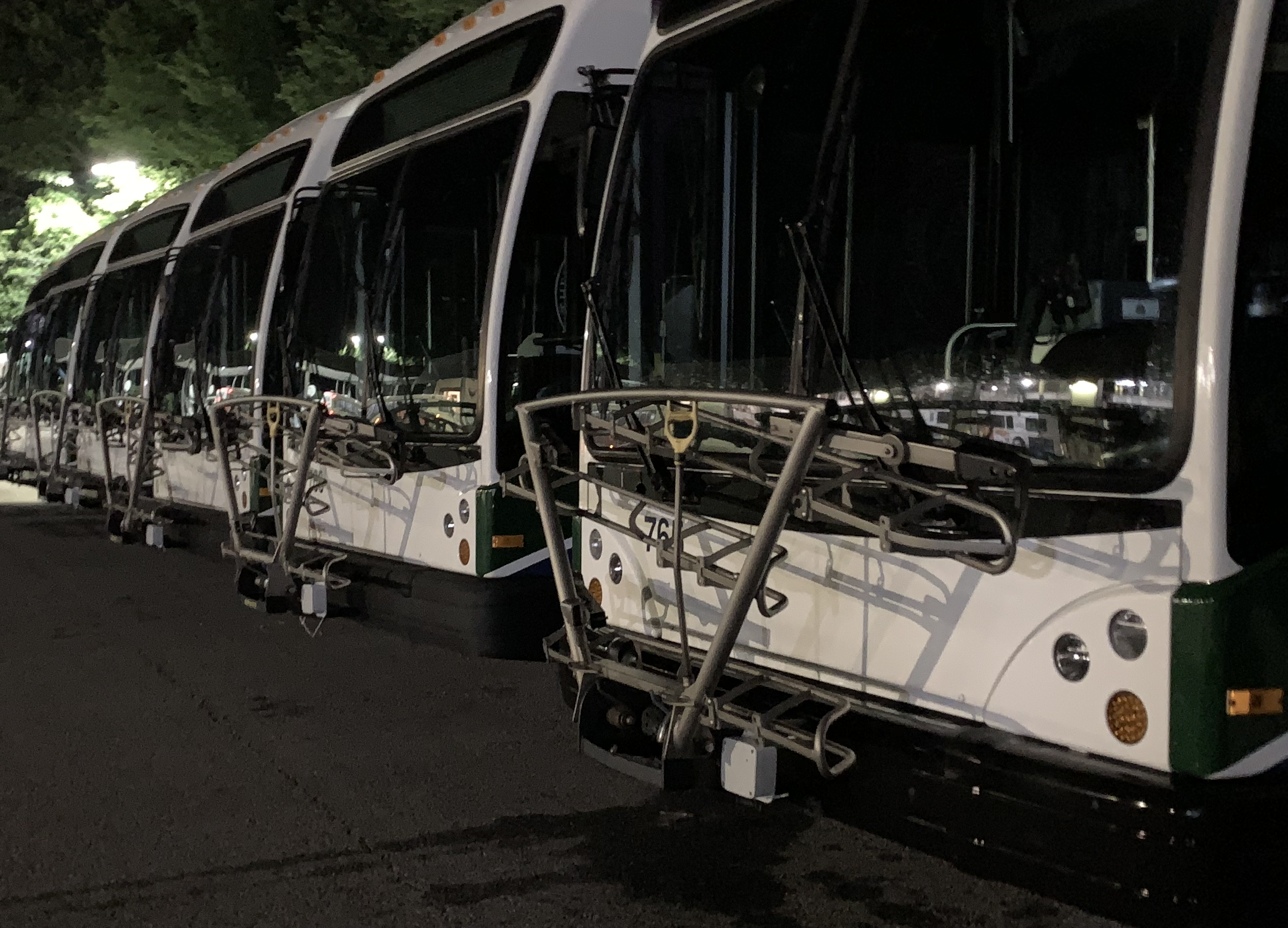}
    \caption{Temperature Sensors Mounted on City Buses}
    \label{fig:sensor-setup}
\end{figure}

Figure \ref{fig:sensor-setup} shows the picture of a temperature sensor setup mounted on public transportation buses in Athens, GA. The sensor setup was assembled in-house by our project team. It consists of an Arduino microcontroller board, DS18B20 1-wire digital temperature sensors with 0.5C accuracy, a low-power GPS sensor, and lithium-ion batteries in a shielded setting. The cost of each sensor setup is approximately \$120. Apart from the maintenance costs, assembling and mounting each sensor setup requires approximately four man-hours. 

\subsection{Mobile Agent Selection Problem}
As stated in the introduction, budgetary and human resource constraints often limit the number of sensor kits that can be mounted in a city's transportation system. On the other hand, the subset of buses that are selected for carrying the sensors has a significant impact on the spatio-temporal coverage of the drive-by sensing platform. 

In order to demonstrate the importance of the bus selection platform, let us consider the ACC public transit system. This system consists of 20 city buses covering the Athens city area (310 km$^2$). And the whole region is modeled as a rectangular grid with 90 by 90 meters square cells. Let us consider a 5-hour time window between 9:00 AM and 2:00 PM on 10-02-2018 (a typical weekday).

\begin{figure}[htbp]
    \centering
    \includegraphics[width=0.85\columnwidth]{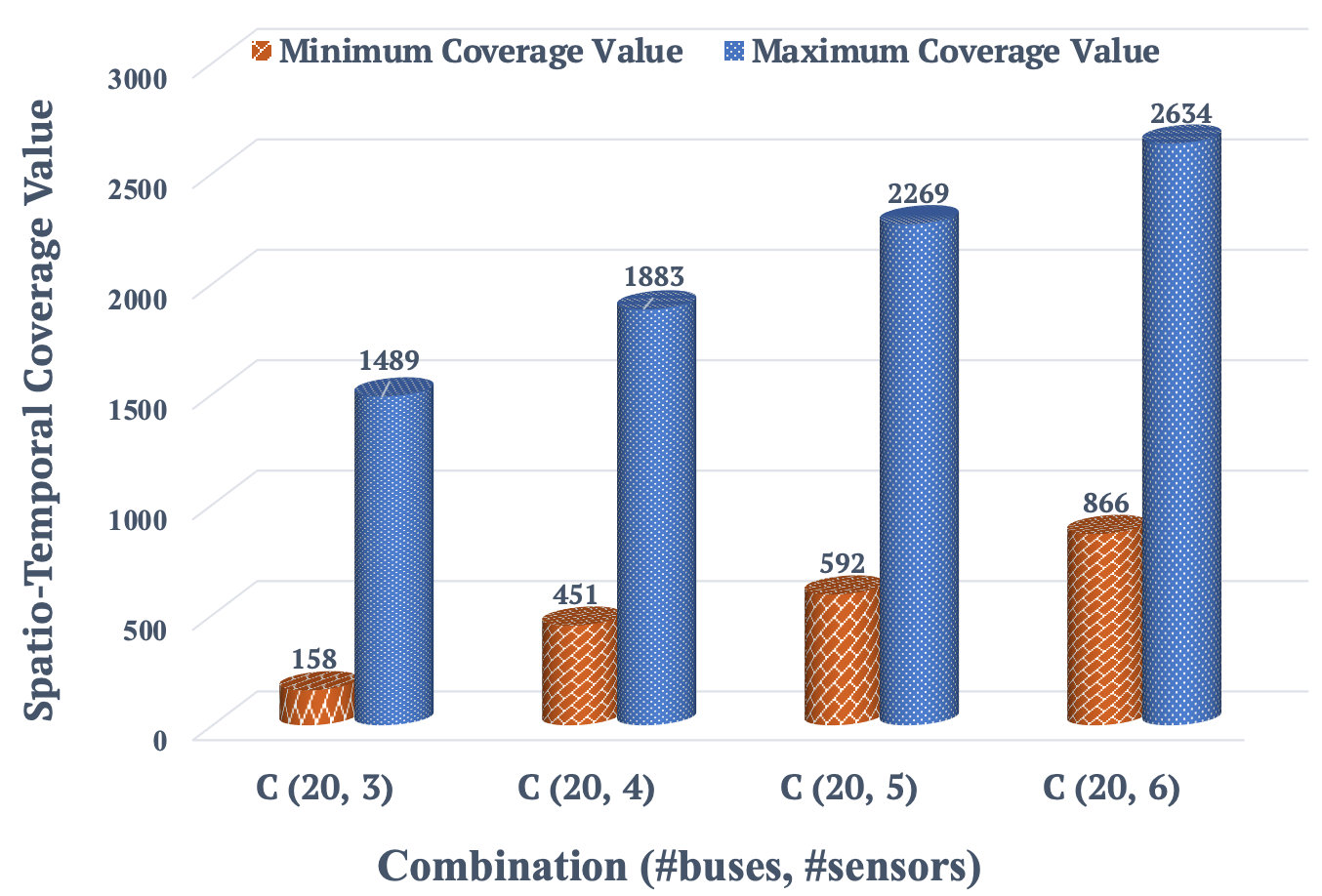}
    \caption{The Worst and The Best Sensing Coverages}
    \label{fig:sensing-coverage}
\end{figure}

Figure \ref{fig:sensing-coverage} shows the  spatio-temporal sensing coverages of the \textit{best} and \textit{worst} possible bus selections when the number of available sensor kits were 3, 4, 5, and 6 respectively. We define spatio-temporal coverage as follows: if a given cell has at least one sensor reading from the region within its boundary (i.e., at least one sensor-carrying bus passed through the cell) in at least $\lambda\%$ of all non-overlapping 1-hour time slots within the 5-hour time window (10:00 AM to 11:00 AM, 11:00 AM to 12:00 PM and so on), its temporal coverage is $\lambda$. The spatio-temporal coverage is the average of the temporal coverages of all the cells in the grid. Note that the spatio-temporal coverage has a direct bearing on the quality of the resultant heat map because it represents the completeness of the underlying data (i.e., if the spatio-temporal coverage is low, it implies a higher percentage of missing data values and vice versa).  

Multiple studies have focused on enhancing the sensing coverage in mobile crowdsensing. These studies assume that all the participants are already equipped with the required sensing devices, and they investigate various approaches to distribute sensing tasks while minimizing recruitment costs.  

Guo et al. \cite{guo2016activecrowd} propose a worker selection approach under two situations: either based on the intentional movement of sensing agents for time-sensitive tasks or based on their unintentional movement for tasks which are not time-sensitive. They show how their proposed algorithm outperforms the previous approaches like that of discussed in \cite{engelbrecht2014particle} as a particle swarm optimization solution. In another study, Campioni et al. \cite{campioni2018improved} analyze recruitment algorithms aimed at selecting participants within a crowdsensing network in a way that the most sensing data is obtained for the lowest possible cost. He et al. \cite{he2015high} present a new participant recruitment strategy for drive-by sensing by predicting the future trajectory of participants. Their proposed algorithms show some improvement in terms of crowdsensing coverage.

In another study, Yi et al. \cite{yi2017fast} propose a fast algorithm for vehicle participant recruitment problem, which achieves a linear-time complexity at the sacrifice of a slightly lower sensing quality. In a separate study, Wang et al. \cite{wang2018maximizing} propose a system model based on the predictable trajectory of public transports through a cloud management platform that interacts with static base stations for distributing the sensing tasks. This research, like the other studies discussed in this section, assumes that all the vehicles are equipped with the required sensors and receive a reward per each sensing task.

\section{MODELING THE MOBILE AGENT SELECTION PROBLEM}
The overall goal of this model is to maximize spatio-temporal coverage of the data set collected through drive-by sensing while taking the relative importance of hotspot locations into account. Therefore, this model focus on selecting an optimal subset of buses in a way to consider the requirements mentioned above. For this purpose, we assume that the trajectory data of the buses are available. In other terms, the routes that each bus traverse are known. So, by using the GPS data and the timestamps associated with them, we can estimate the location of each bus at a particular point in time.

\begin{figure}[htbp]
    \centering
    \includegraphics[width=0.5\columnwidth]{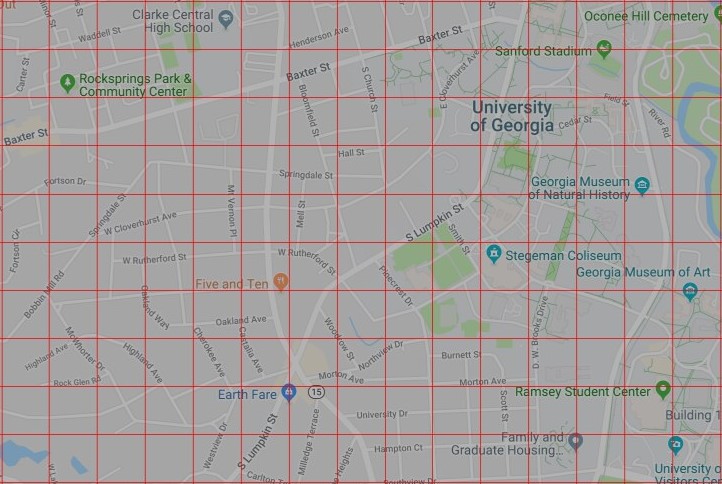}
    \caption{A Sample Grid Representation}
    \label{fig:grid}
\end{figure}

To formulate the mobile agent selection problem, we model the area as a grid of square cells, as shown in Figure\ref{fig:grid}. Each cell is characterized by the GPS coordinate of its four corners. The dimension of each cell is a configurable parameter and represents the spatial granularity of the sensing. We define matrix A, where an arbitrary cell of the grid is represented as $a_{ij}$:

$$
A = 
\begin{bmatrix} 
a_{11} & ... & a_{1n}\\
:      &     & :     \\
a_{m1} & ... & a_{mn}
\end{bmatrix}
$$

The relative importance of hotspot areas in a city is represented by the weights assigned to their corresponding cells in the grid structure. Therefore, matrix W is defined where each grid cell is associated with a weight:

$$W = 
\begin{bmatrix} 
w_{11} & ... & w_{1n}\\
:      &     & :     \\
w_{m1} & ... & w_{mn}
\end{bmatrix}
$$

In our design, time is modeled as a vector of $T = \{t_1, t_2,..., t_l\}$ where each $t_k$ is a time slot with configurable duration. The sum of these time slots is 24 hours, and the duration of each slot represents the granularity along the temporal dimension. For example, if we need to have a reading of an environmental feature every 30 minutes, each $t_k$ would denote a 30-minute time slot. 

The set of $B = \{b_1, b_2, ..., b_p\}$ represents all the buses available in the city where each $b_\lambda$ represents an individual bus. If a bus $b_\lambda$ carries a sensor (i.e., it is selected for sensor deployment), it can obtain a reading from the cell $a_{ij}$ in the time slot $t_k$ if and only if $b_\lambda$ is present within $a_{ij}$\textquotesingle s boundaries for at least some duration of time slot $t_k$ (i.e., $b_\lambda$ has traversed through $a_{ij}$ in time slot $t_k$). Please note that a bus can traverse through multiple cells during a time slot. Also, multiple buses can traverse through a same cell during the same time slot (in which case, we obtain duplicate values). Considering the limited Number of Sensors (NS), we define: $$BS = \{BS_1, BS_2, ..., BS_q\} $$ as the set of all possible bus combinations, where each $BS_i$ is a set of buses ($BS_i \subseteq B$) and the size of each of these set is less than or equal to the number of available sensors ($|BS_i| \leq NS$). For instance $BS_1$ can be represented as:
$$BS_1=\{b_5, b_{18} , b_{24}\}  $$

\subsubsection*{\textbf{OBJECTIVE FUNCTION}}
In this section, we define an objective function for bus selection that reflects the overall goal of maximizing the spatiotemporal coverage. For this purpose, let's suppose that the Selected Bus Set of $SBS^* = \{b_l, b_k, b_p\}$ represents the set of 3 buses which are selected for sensor deployment, such that $SBS \subseteq B$ and $|SBS| \leq NS  $. Having laid out the model, we now define the Coverage Value (CV) of $BS$ with respect to a cell $a_{ij}$ at a time slot $t_k$ as follows:

\begin{equation}
 CV(BS_x,a_{ij},t_k) =
 \begin{cases}
      w_{ij}^{t_k},\text{if}\{\exists b_i \in BS_x | b_i \ is  \ in \ a_{ij} \ at \ t_k\}  \\
      0, \text{otherwise}
    \end{cases}
\end{equation}

In other words, CV determines whether at least one of the buses in a set sensed the given cell at the given time slot or not. If the condition is true, the bus set gains the coverage value associated with that location, which is equal to the cell's weight. Otherwise, the set gains no coverage value for that specific time slot.

In the next step towards our objective function, we define the Cumulative Coverage Value (CCV) of a bus set as:

\begin{equation}
 CCV(BS_x) =\sum_{t_k \in T} \sum_{\forall a_{ij} \in A} CV(BS_x,a_{ij},t_k)
\end{equation}

This measure calculates the aggregated coverage value of a bus set in all the time slots during a day while eliminating the duplicate values. In other terms, if more than one bus in a set covers a grid cell in the same time slot, the weight associated with that cell will be added to the CCV only once.

Finally, the Selected Bus Set ($SBS^*$) will be the bus set that its CCV is higher than all other possible bus combinations. If more than one set achieves the same maximum CCV, the set in which its minimum CV in all the time slots is higher than that of the other sets, will be chosen (it denotes the set with better spatial coverage in each single time slot). Therefore, our objective function is defined as follows:

\begin{equation}
SBS^* =BS_x,  \text{if} \ \ (CCV(BS_x) > \bigwedge\limits_{BS_i-\{BS_x\}} CCV(BS))
\end{equation}

In short, the primary motivation is to minimize the redundant values in both space and time by selecting the best subset of our mobile agents.

\subsubsection*{\textbf{ILLUSTRATIVE EXAMPLE}}
\begin{figure}[htbp]
    \centering
    \includegraphics[width=2.7in]{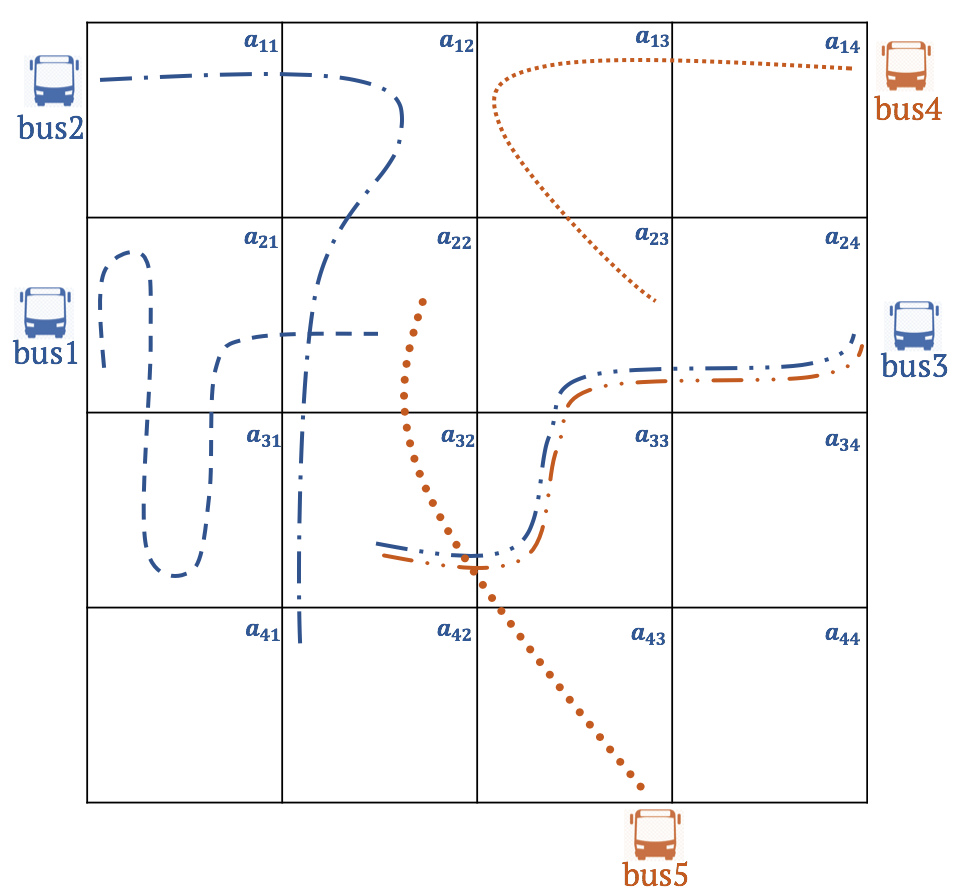} 
    \caption{An Example of a Bus Selection Coverage in One Time Slot}
    \label{fig:busselection}
\end{figure}

In order to better understand the definitions mentioned above, Figure \ref{fig:busselection} shows a sample grid with 16 cells and no hotspot ($w_{ij}$=1). The routes that each bus passed during a time slot is depicted using the dotted lines. Let\textquotesingle s suppose that there are two bus selections named $BS_1$ and $BS_2$, where:
$$BS_1 = \{bus_1, bus_2, bus_3\}$$
$$BS_2 = \{bus_3, bus_4, bus_5\}$$

\begin{table}[htbp]
\centering
\caption{Calculating Bus Coverage Value at $t_l$}
\includegraphics[width=2.9in]{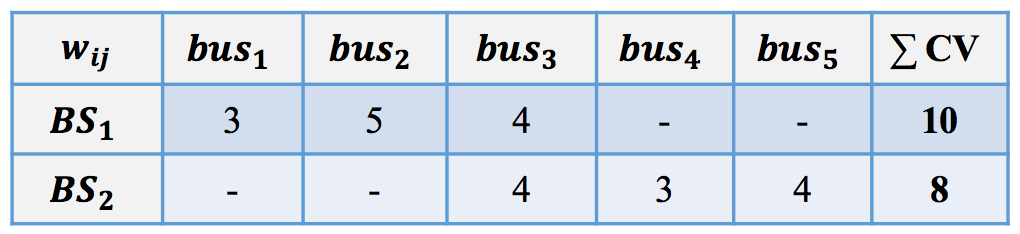}
\label{tbl:BusSelectionReward}
\end{table}

\begin{figure}[htbp]
    \centering
    \includegraphics[width=1\columnwidth]{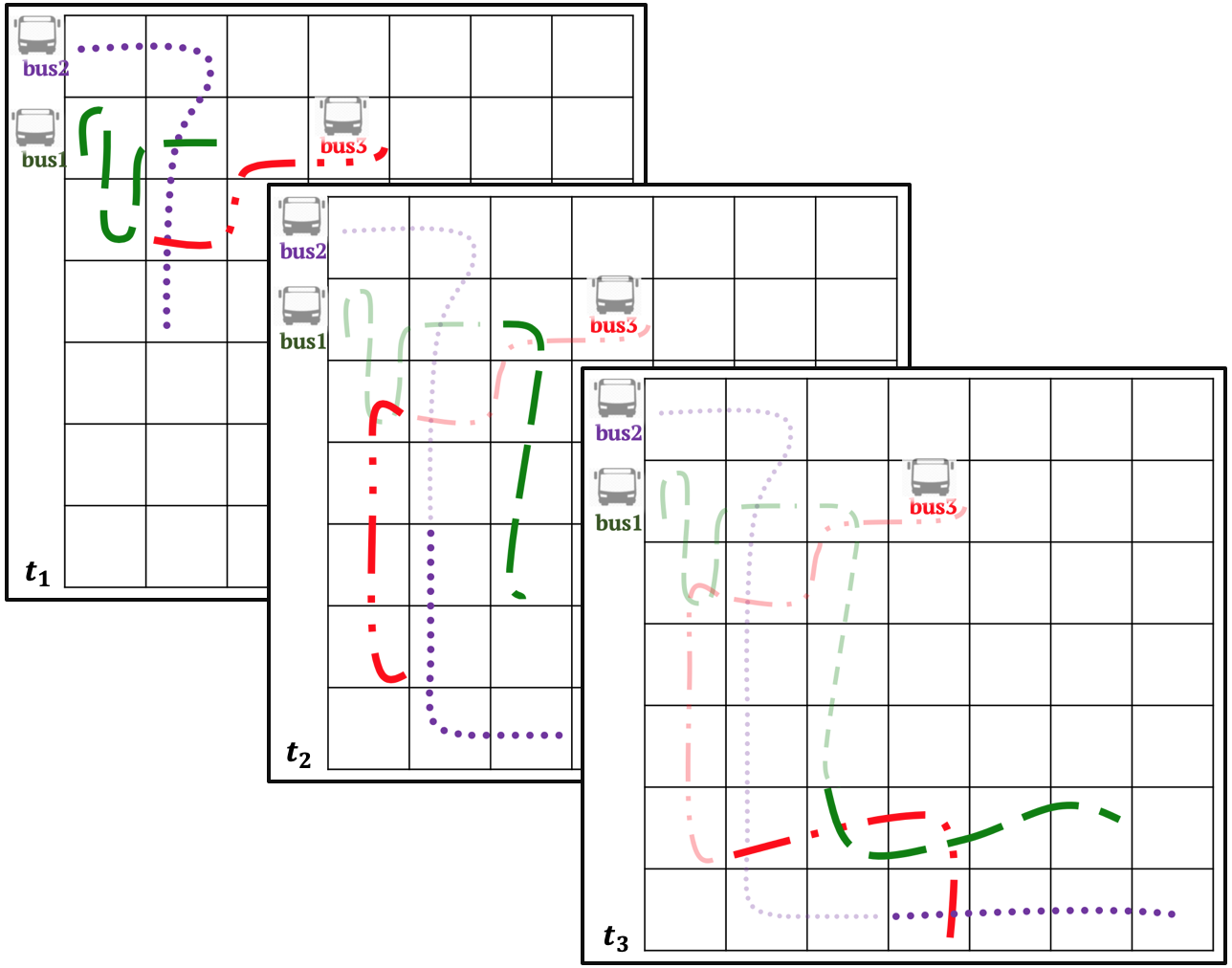} 
    \caption{An Example of a Bus Selection ($BS_1$) Coverage in Three Consecutive Time Slots}
    \label{fig:s1ts}
\end{figure}

Although $bus_3$ is selected in both sets, the other two buses are different. Table \ref{tbl:BusSelectionReward} represents the number of cells passed by each bus during a given time slot. For instance, $bus_1$ passed three different cells ($a_{21}$, $a_{31}$, and $a_{22}$); therefore, it gets the coverage value of 3. Similarly, the coverage value for other buses is calculated. The last column represents the sum of the coverage values while excluding the duplicates. Therefore, although the actual sum of the values in the first row equals 12, eliminating the duplicates reduces it to 10. As depicted in Figure \ref{fig:busselection}, we can see that the two cells of $a_{22}$, $a_{32}$ are covered twice. Thus, the first bus set as a whole, gained a coverage value of 10 in this time period. 

In the next step, we want to continue with the same example in Figure\ref{fig:busselection} for three consecutive time slots. In Figure\ref{fig:s1ts} the grid on the back corresponds to the same bus set of $BS_1$ which we saw in Figure\ref{fig:busselection}. Considering that during the first time slot, $BS_1$ covered 10 different cells, this selection gains the coverage values of 10 in $t_1$. During the second time slot, the three buses continued their routes and sensed 12 different cells. Although some of the cells were already sensed during $t_1$, these cells are counted again in $t_2$, because we only exclude the overlaps within the same time slot. Therefore, $BS_1$ gets the coverage value of 12 in $t_2$. Following the same logic, $BS_1$ gains the coverage value of 9 during $t_3$. These coverage values correspond to the first row of Table \ref{tbl:TotalRewardperTime}.

\begin{table}[htbp]
\centering
\caption{Total Sensed Cells Per Each Sensing Period}
\includegraphics[width=1.8in]{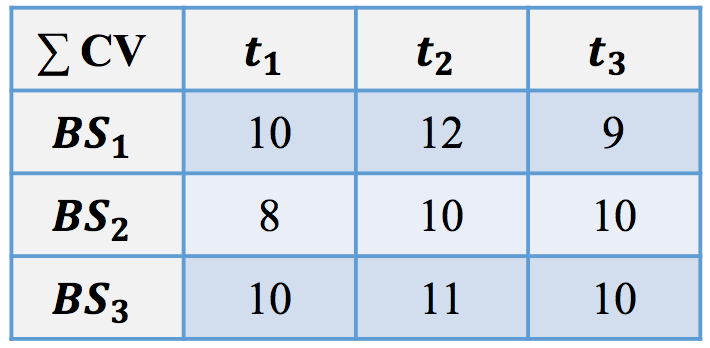}
\label{tbl:TotalRewardperTime}
\end{table}

In the next step, we generate the Total Coverage Value for each bus set during the whole time period. The first column in Table \ref{tbl:TotalSensingPeriod} represents the CCV for each $BS_i$. The values of this column are simply the summation of the values in each row of Table \ref{tbl:TotalRewardperTime}. The second column of Table \ref{tbl:TotalSensingPeriod} shows the minimum value of each row of Table \ref{tbl:TotalRewardperTime}. In other terms, this column shows the minimum coverage values that each bus set earned during each time slot.

\begin{table}[htbp]
\centering
\caption{Total Sensing Coverage Value for Each Bus Selection During the Whole Time Period}
\includegraphics[width=2in]{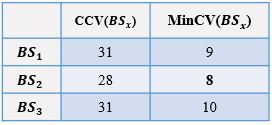}
\label{tbl:TotalSensingPeriod}
\end{table}

\begin{figure}[htbp]
    \centering
    \includegraphics[width=2.7in]{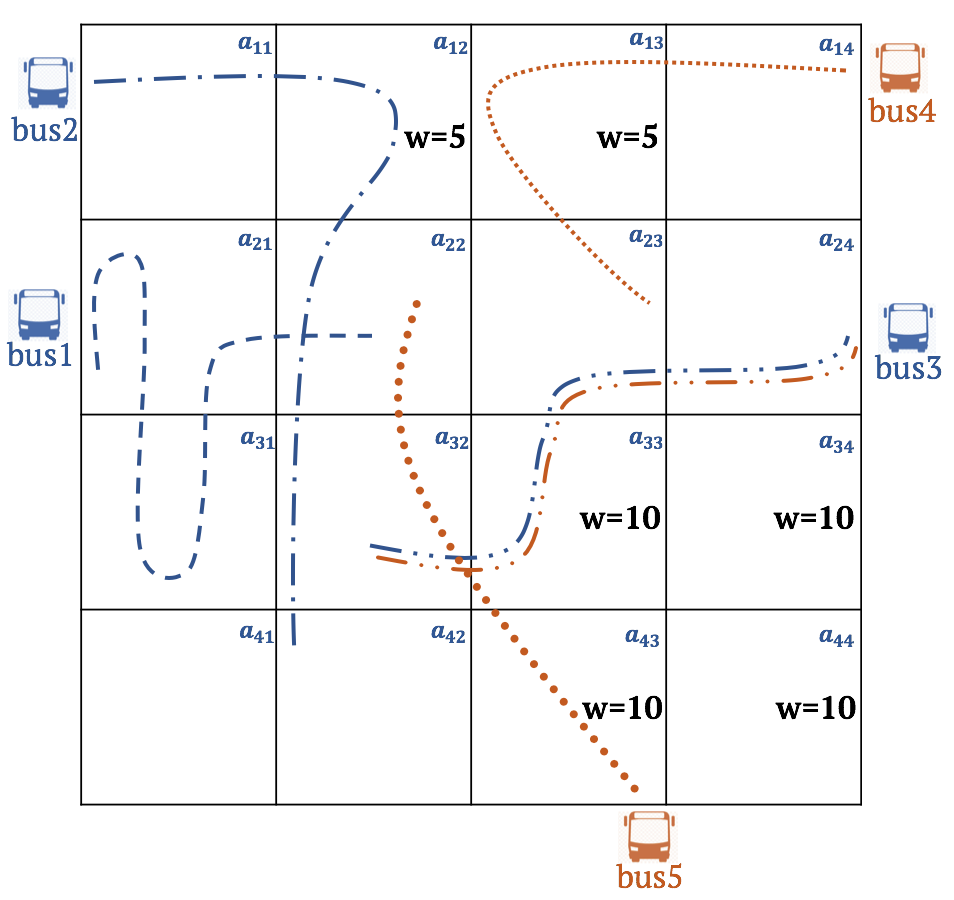} 
    \caption{An Example of a Grid with hotspots of Different Weights}
    \label{fig:DifferentWeights}
\end{figure}

To better understand how the different weights of hotspot locations can affect the sensing coverage values, Figure \ref{fig:DifferentWeights} depicts the previous example with $BS_1$ and $BS_2$ while the grid cells have different weights. Accordingly, Table \ref{tbl:W-Reward} shows the updated coverage values of these two bus sets  at $t_1$.

\begin{table}[htbp!]
\centering
\caption{Calculating Bus Selection Coverage Value at $t_l$ with Hotspots}
\includegraphics[width=2.8in]{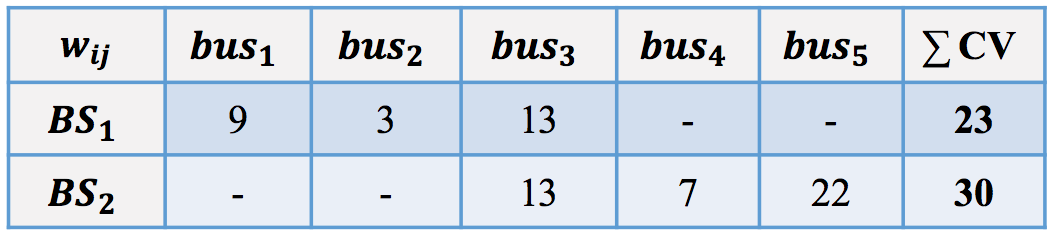}
\label{tbl:W-Reward}
\end{table}

\section{SIMPLE APPROACHES AND THEIR LIMITATIONS}
\subsection{NAIVE APPROACH}
The simplest approach to solve the problem is to mount sensors on a randomly selected set of buses. Since this approach does not consider any requirement with respect to spatiotemporal coverage or the hotspot locations, it will often lead to a poor selection of buses. Furthermore, the variance of CCV among different runs of the algorithm will be very high.

\subsection{EXHAUSTIVE APPROACH}
The other approach which considers our discussed objective function is the exhaustive method. In this approach, all the possible combinations of \textit{n} buses taken \textit{r} at a time, where r is equal to the number of sensors (NS), is computed. Then, the bus combination with the highest CCV will be chosen. 

In this algorithm, we first need to create the grid structure based on the given size for each cell by using the latitude and longitude of the area. Next, the algorithm generates the matrix W, where the weights associated with each grid are provided by domain scientists based on the target phenomena to be monitored. Besides, this algorithm creates all the possible bus selections with X different buses where X is equal to or less than the number of sensors. Furthermore, it calculates the set of time slots within the total sensing period. Given these data, the main function chooses the most optimal bus set with the highest spatial and temporal coverage.

This algorithm calls two other functions. The first function, which is called $CCV\_Calculation$, determines the cumulative coverage value earned by each given bus set by looping through the set of mobile agents, the cells within the grid structure, and the weights associated with each grid cell. Furthermore, it calculates the minimum coverage value during different time slots for each grid cell. The second function, called $SBS$, chooses the best selection by applying the objective function. In other terms, it finds the bus selection with the highest CCV.

Although this method is computationally expensive (its runtime grows factorially in terms of the number of bus combinations), it is guaranteed to choose the best possible bus combination where the CCV is higher than all other bus sets.

Considering that the exhaustive approach calculates all the r--combinations of the set of mobile agents where r is the limited number of sensors, running the algorithm for large data sets leads to extremely long processing time. There are many applications where the sensing parameters, such as the coverage values associated with each hotspot, changes quickly. Thus, we have to unmount and mount our sensors on a new subset of buses to monitor the target environmental features in a dynamic setting. For instance, a football game may necessitate extra surveillance coverage. Therefore, there should be mechanisms to select an optimal subset of buses to mount surveillance cameras for monitoring the areas around the stadium for that particular day. As a result, there is a need for utility-aware approaches with a fast decision process to choose an optimal subset of public transportation vehicles to cover the target areas.

To provide a better understanding of the scale of real-world applications, Table \ref{tbl:buswolrd} provides the number of buses in some selected cities around the world \cite{Atlanta, NewYork, Washington, LosAngeles, Karachi, Beijing}. It also represents the number of different bus combinations if 5\%, 10\%, or 20\% of the buses are supposed to be selected. For instance, there are 639 buses in Atlanta. If we want to select 32 buses out of 639 which traverse around this city, we need to calculate the CCV of around 1.03E+54 different bus selections.

\begin{table*}[htbp]
\centering
\caption{Combinations of Different Bus Selection in Selected Cities}
\includegraphics[width=5in]{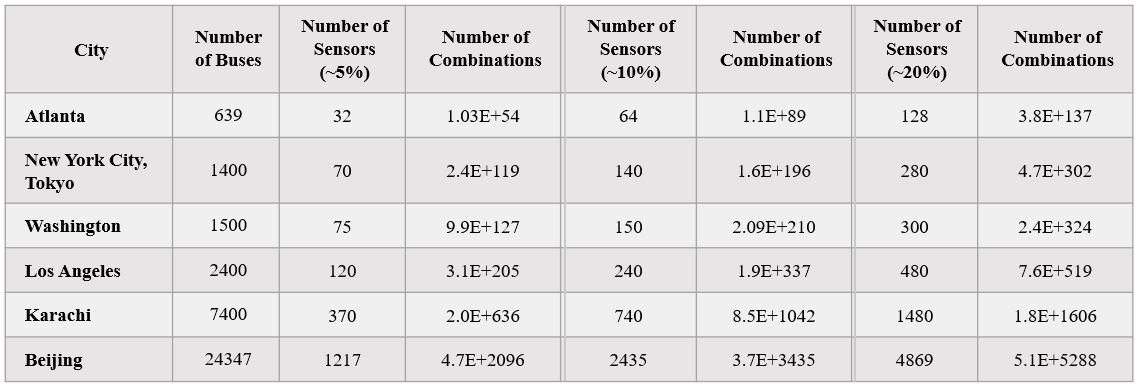}
\label{tbl:buswolrd}
\end{table*}

\section{UTILITY-AWARE APPROACHES}

To resolve the limitations associated with the simple approaches, in this section, we propose our utility-aware sensing approaches, which can be leveraged in various sensing frameworks where selecting a subset of vehicles is required.

\subsection{HOTSPOT-BASED APPROACH}

The hotspot-based approach is designed based on the relative importance of some areas in an urban region. The importance of a particular area is indicated by a weight ($ w_{ij}$) assigned to the corresponding grid cell, while the default weight of each grid cell is one.

In this approach, instead of running the aforementioned exhaustive algorithm on all the grid cells, we only consider hotspot cells, i.e., the cells that correspond to areas with higher importance levels as indicated by their respective weight values. The threshold of these weight values for a cell to be considered a hotspot is a configuration parameter and is specified at the time of running the algorithm. For instance, the locations with a high variation in temperature or the areas with high population density can be configured as hotspot locations.

On the other hand, this approach excludes the buses that do not pass through any hotspot location. Excluding buses that do not pass through hotspots significantly reduces the number of bus combinations that need to be considered, thus letting the algorithm to perform much faster. 

The hotspot-based approach can be either used standalone or act as the initial step of our genetic algorithm, which is more rewarding, but slightly slower. Our utility-aware genetic algorithm, which is founded on top of this hotspot-based algorithm, will be discussed in the following section.

\subsection{UTILITY-AWARE GENETIC ALGORITHM}
Genetic algorithms are very suitable candidates to solve optimization problems where finding the best possible answer is very computationally-expensive. Considering that our hotspot-based approach only focuses on the hotspot locations to make the decision, we propose a customized genetic algorithm that considers the buses with high spatiotemporal coverage while they may not have covered any hotspot area. This algorithm uses the output of our hotspot-based algorithm as its input and provides the chance of exploring non-hotspot areas.

\begin{figure}[htbp]
    \centering
    \includegraphics[width=1\columnwidth]{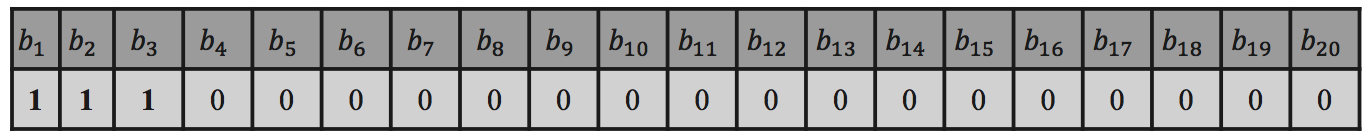} 
    \caption{A Sample Chromosome Representation}
    \label{fig:chromosomeRepresentation}
\end{figure}

In our genetic algorithm, chromosome representation is used to encode the candidate buses. For instance, if we wanted to select three buses out of twenty, we would have a chromosome representation like that of Figure \ref{fig:chromosomeRepresentation}, where there are precisely three 1s. This condition enforces the algorithm to select the number of buses proportional to the number of sensors.

\begin{figure}[htbp]
    \centering
    \includegraphics[width=1\columnwidth]{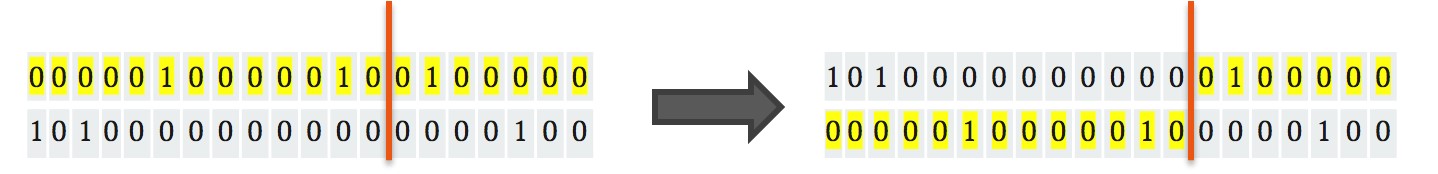} 
    \caption{A Sample Crossover Operation}
    \label{fig:crossover}
\end{figure}

The crossover (recombination) operator combines the genetic representation of two parents to create a new generation. In our design, the algorithm randomly selects a single crossover point in the chromosome representation of the two parents and recombines them like the example shown in Figure \ref{fig:crossover}. Thus, the bits to the right of the selected crossover point will be swapped between the two parent chromosomes so as to generate two new child chromosomes.

\begin{figure}[htbp]
    \centering
    \includegraphics[width=1\columnwidth]{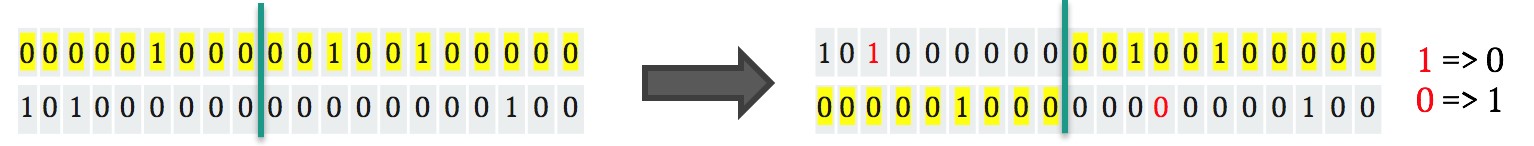} 
    \caption{A Sample Mutation Operation}
    \label{fig:mutation}
\end{figure}

After each crossover operation, our algorithm checks whether the number of 1s in each new chromosome still corresponds to the number of sensors or not. If the condition is not met, the mutation operation will be used to randomly flip bits in each child until the condition is satisfied. Figure \ref{fig:mutation} depicts a crossover operation that invalidates the condition mentioned earlier; therefore, the mutation operation comes into the picture to solve the inconsistency.

Considering that the fitness function of our genetic algorithm is the same as the objective function of our exhaustive algorithm, in the selection and replacement phase, our initial population is chosen from the results of our hotspot-based approach. In other terms, we first calculate the actual CCV of the bus sets generated by the hotspot-based approach and sort them. Next, based on the experimental setup, we select our initial chromosome population from the sorted list. Then, in each iteration, based on the coverage values, we discard the worst 20\% of the population and replace them with new children generated from the parents coming from the top 20\% of the population. 

\begin{figure}[htbp]
    \centering
    \includegraphics[width=1\columnwidth]{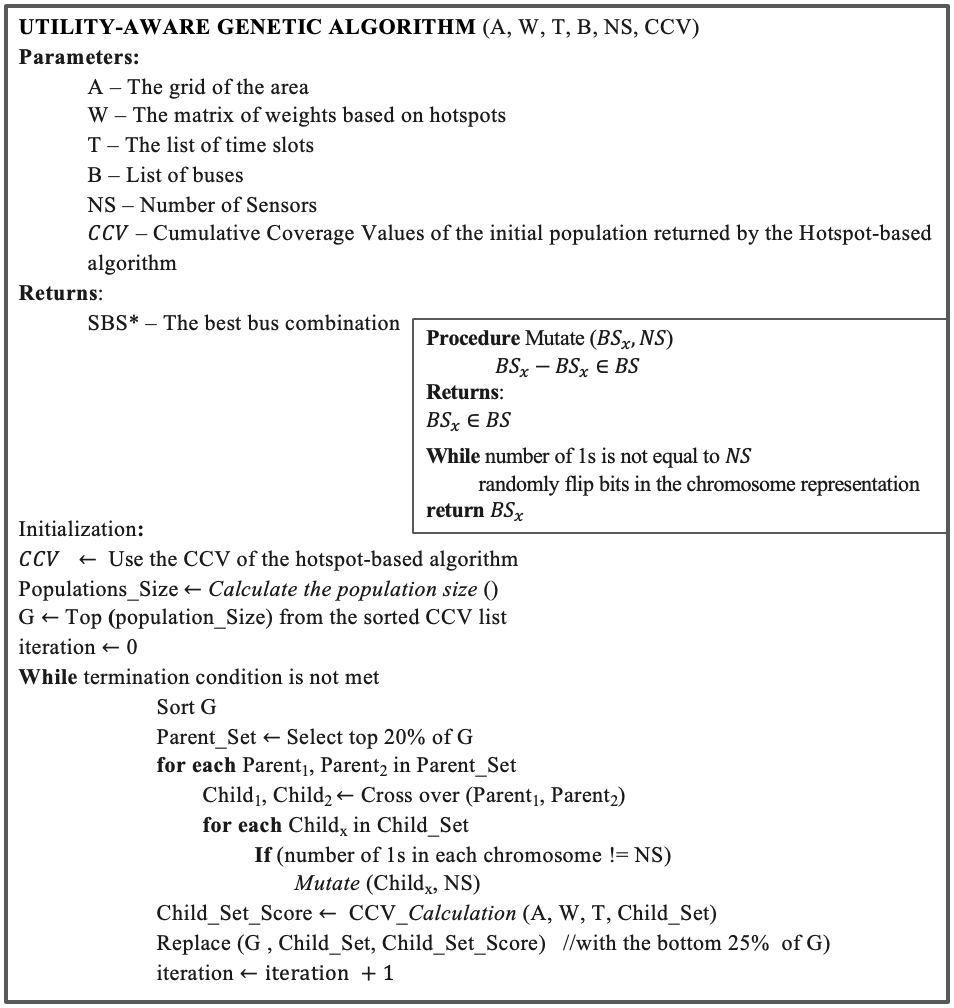} 
    \caption{Pseudocode of the Utility-Aware Genetic Algorithm}
    \label{fig:GA}
\end{figure}

As shown in the pseudocode of our utility-aware genetic algorithm in Figure\ref{fig:GA}, the algorithm initializes with the population of buses chosen by the hotspot-based algorithm. Next, it performs the crossover operation based on the replacement criteria (and mutation, if required). The crossover and mutation steps provide the algorithm with the chance to consider new bus sets that were disregarded by the hotspot-based algorithm. Then, the coverage values for the new generation will be recalculated and sorted. The algorithm continues until one of the two termination conditions is satisfied; it stops once the results start converging; otherwise, it terminates by reaching the iteration threshold.

Although the genetic algorithm is much more efficient in terms of computation time compared to the exhaustive algorithm, its core idea of analyzing each bus combination as a whole, is the same. In the following section, we present a utility-aware redundancy minimization approach, that unlike the previous algorithms, selects one bus at a time.

\subsection{UTILITY-AWARE REDUNDANCY MINIMIZATION ALGORITHM}
Our redundancy minimization algorithm is designed as follows: in the first step, the bus that has the highest spatiotemporal coverage and goes through the largest number of hotspots in different time slots is selected. In the next step, the algorithm chooses the second bus with the best coverage while it passes through the most number of remaining hotspots. In other words, in each step, the selection is made in a way to maximize the number of sensed hotspots, excluding the ones which are already covered by the previous mobile agents. This algorithm continues selecting one mobile agent at a time until it reaches the limit imposed by the number of sensors.

\begin{figure}[htbp]
    \centering
    \includegraphics[width=2.5in]{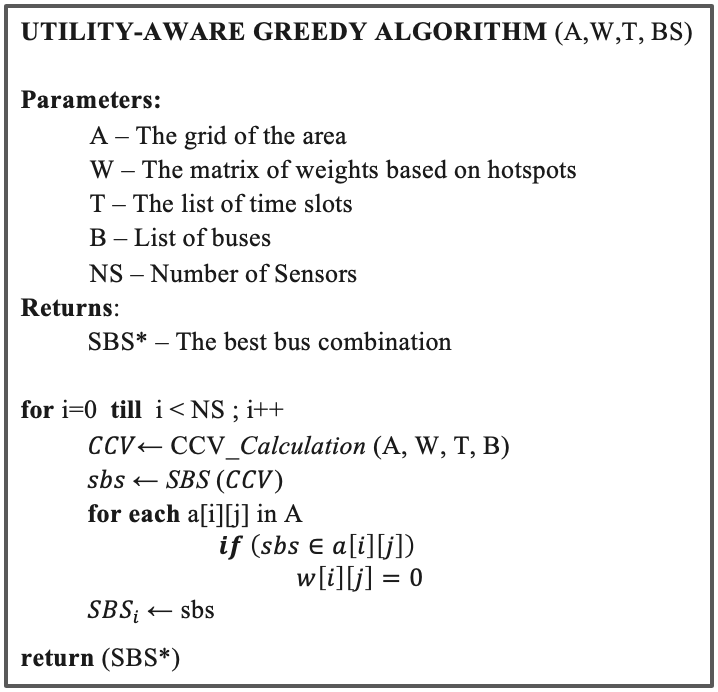} 
    \caption{Pseudocode of the Utility-Aware Redundancy Minimization Algorithm}
    \label{fig:LAKSGREEDY}
\end{figure}

As shown in the pseudocode of our utility-aware spatiotemporal redundancy minimization algorithm in Figure\ref{fig:LAKSGREEDY}, the algorithm starts by choosing the bus with the highest CCV. Upon selecting a bus, the weight associated with all the grid cells where that specific bus covered in each time slot will be changed to zero. Therefore, by excluding the overlaps, the next bus will be selected such that it covers the highest number of remained cells and hotspots. 

In general, greedy strategies, like our redundancy minimization algorithm, are promisingly efficient in analyzing large spatiotemporal data sets \cite{pirsiavash2011globally}. In particular, these approaches become very useful for solving combinatorial optimization problems. Big data analysis necessitates leveraging scalable computational methods that can be used in real-world scenarios where a fast and efficient decision-making process is required.

\section{EXPERIMENTAL EVALUATION}
In this section, we perform a couple of experiments to analyze the spatiotemporal coverage and the computational performance of our proposed algorithms on a real-world data set. In this case study, we focus on the SCOUTS project, the goal of which is to generate hyperlocal heatmaps of the urban areas with high spatiotemporal granularity. The data set includes one-year trajectory data of twenty city buses in the city of Athens, Georgia.

\subsection{EXPERIMENTAL SETUP}

To solve the coverage maximization problem of city buses, we create a grid covering the whole area while each cell corresponds to a 90-meter by 90-meter area on earth. In this study, the hotspot locations and their respective weights are determined by the heatmaps generated from satellite imagery. In other terms, the relative importance of each hotspot to be targeted by drive-by sensing vehicles is defined by analyzing the history of heatmaps generated by Landsat 8 satellite imagery.

We tested our proposed algorithms on 5 hours of bus trajectory data collected every 5 seconds, which in total contains more than 61,000 data points. Furthermore, we located seven different hotspots in the city of Athens, with their weights varying from 2 to 8. It should be mentioned that the hotspot locations provided by the remote sensing experts in this experimental study, covered less than 0.075\% of the whole urban area.

\subsection{RESULTS}
\begin{figure}[htbp]
    \centering
    \includegraphics[width=0.95\columnwidth]{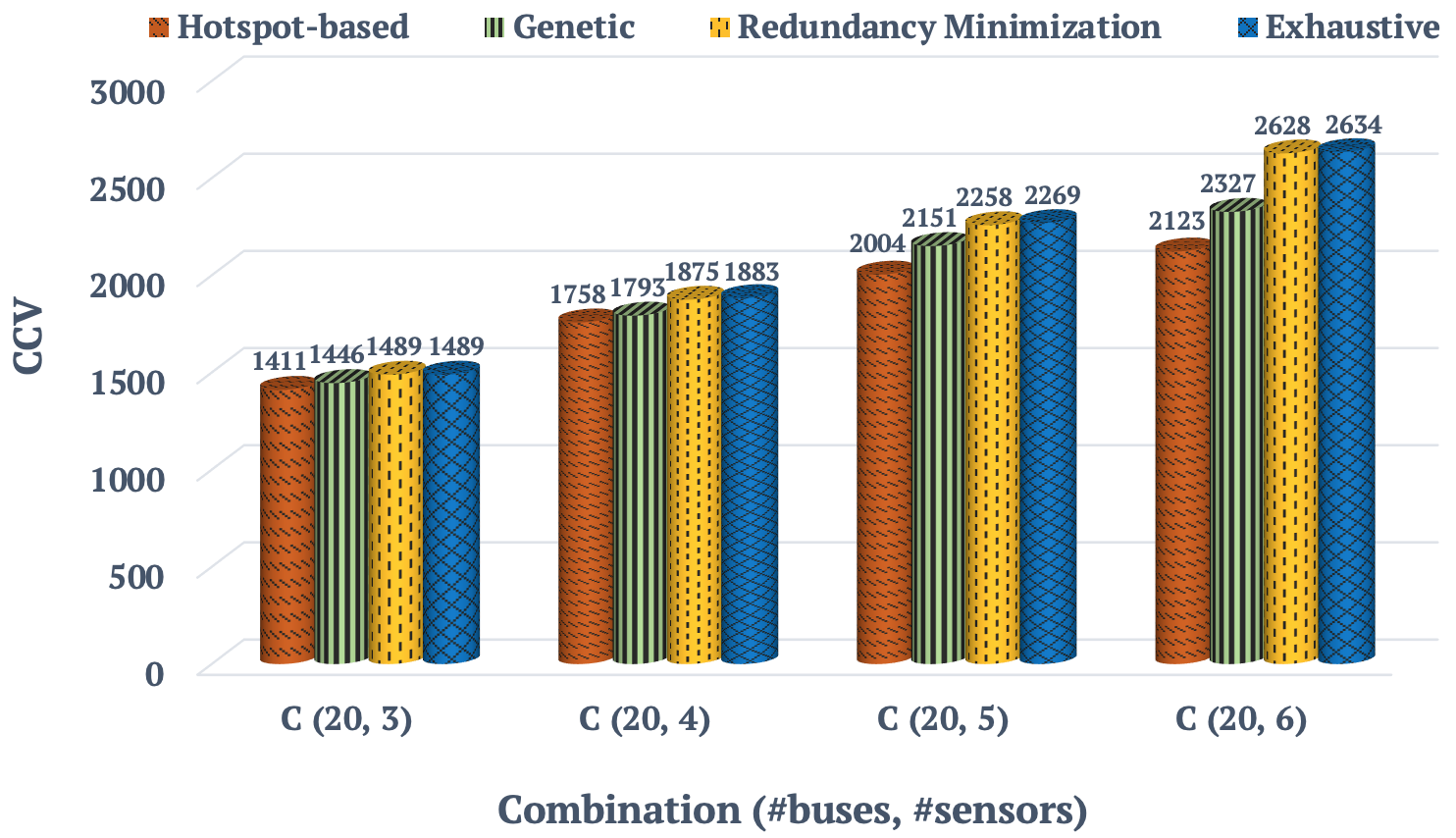} 
    \caption{CCV Comparison for Different Number of Sensors}
    \label{fig:cvvcompare}
\end{figure}

\begin{figure}[htbp]
    \centering
    \includegraphics[width=0.95\columnwidth]{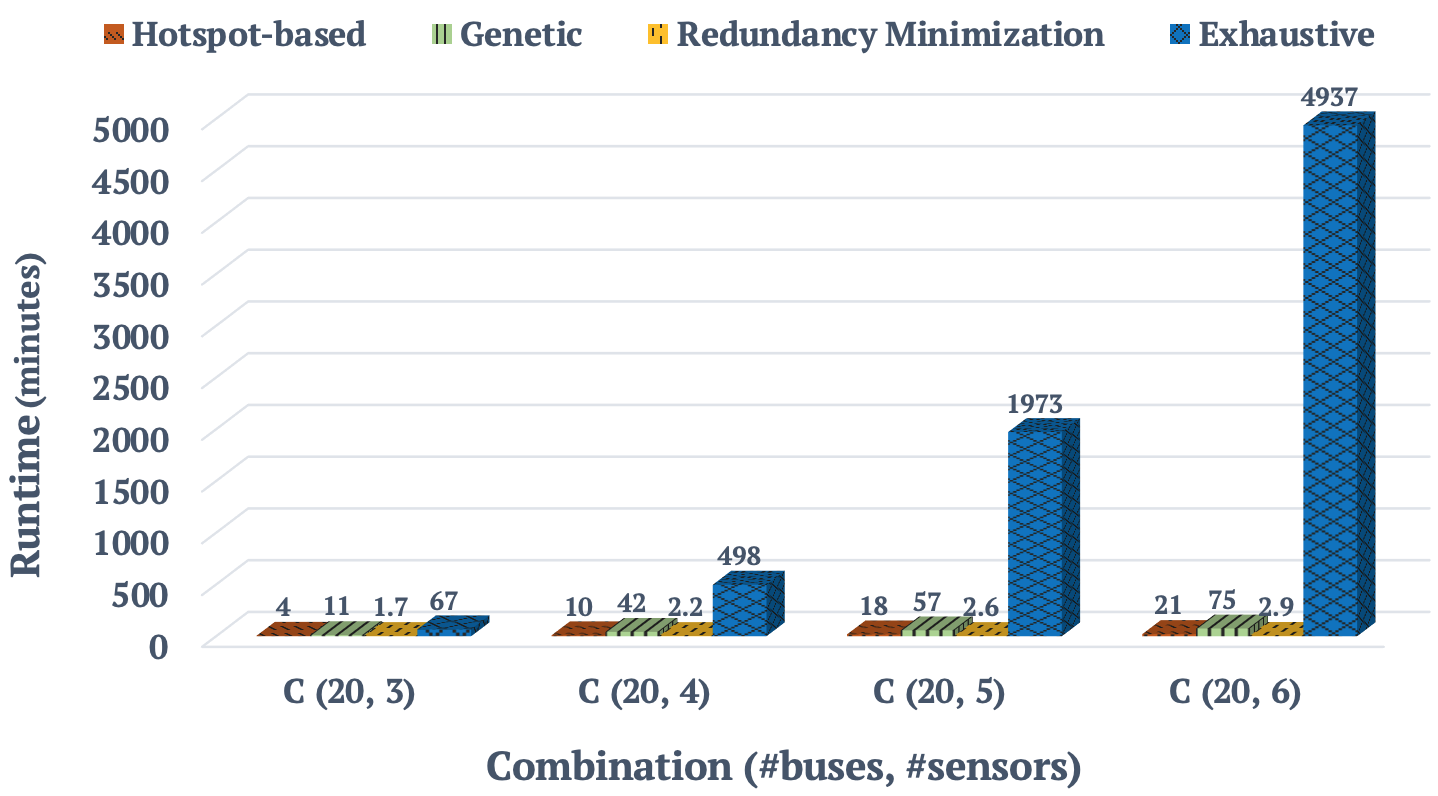} 
    \caption{Runtime Comparison for Different Number of Sensors}
    \label{fig:runtime}
\end{figure}

In order to compare the coverage values along with the runtime performance of our four algorithms, we tested them on a varying number of sensors (i.e., 3, 4, 5, and 6 sensors). Figure \ref{fig:cvvcompare} shows the CCV earned by each algorithm and Figure \ref{fig:runtime} depicts the runtime comparison of these algorithms. We can see that the utility-aware redundancy minimization algorithm consistently gains a CCV very close to the highest possible coverage value, while its runtime is orders of magnitude less than the exhaustive algorithm. In other terms, although the exhaustive algorithm guarantee to choose the bus combination with the highest possible CCV, its substantial computational cost, which grows significantly, makes it impractical for real-world applications. The considerable performance of our redundancy minimization algorithm is followed by the genetic algorithm, and the hotspot-based algorithm, respectively.

\begin{figure}[htbp]
    \centering
    \includegraphics[width=0.95\columnwidth]{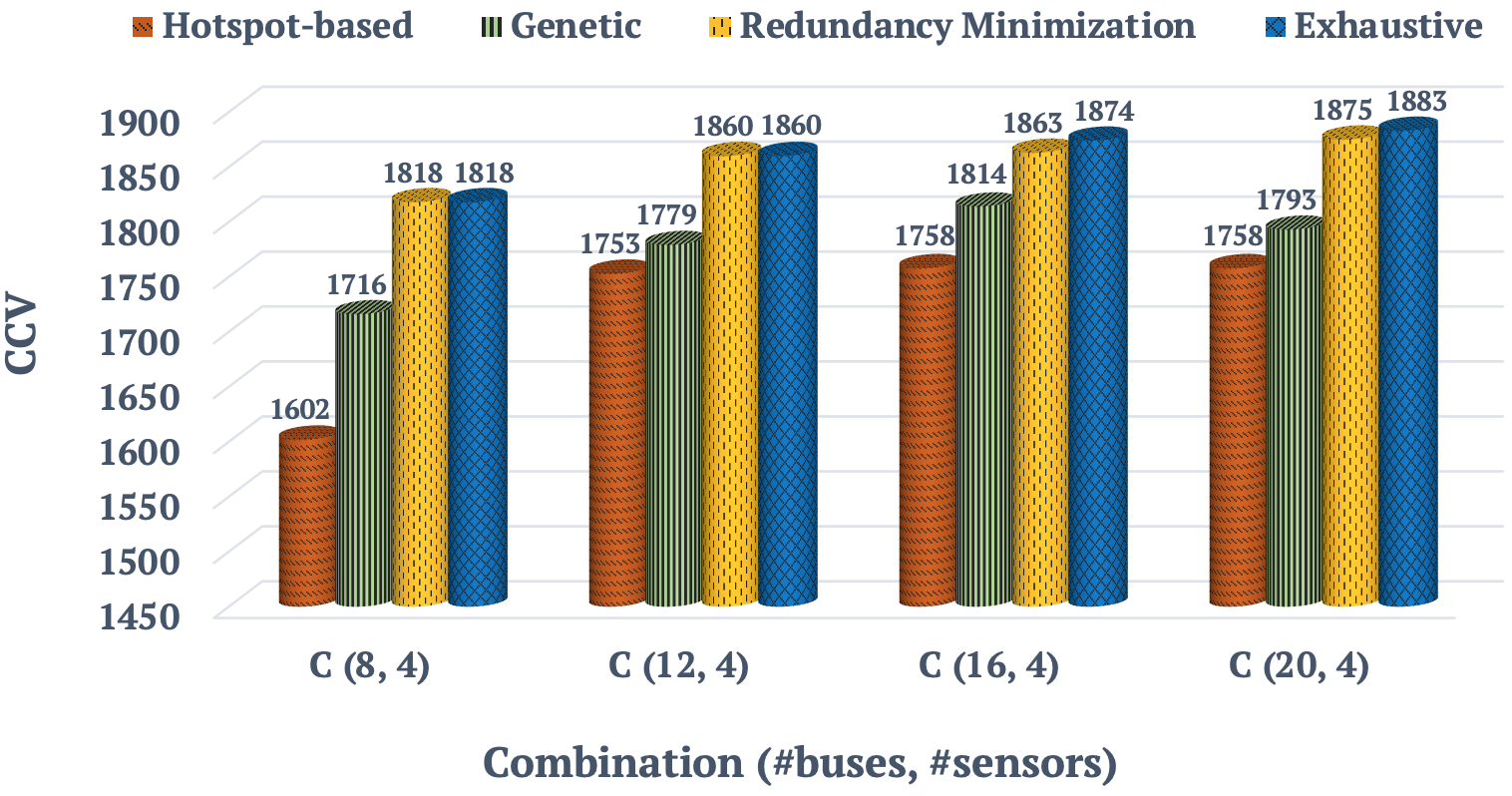} 
    \caption{CCV Comparison for Different Number of Buses}
    \label{fig:z4cvvcompare}
\end{figure}

\begin{figure}[htbp]
    \centering
    \includegraphics[width=0.95\columnwidth]{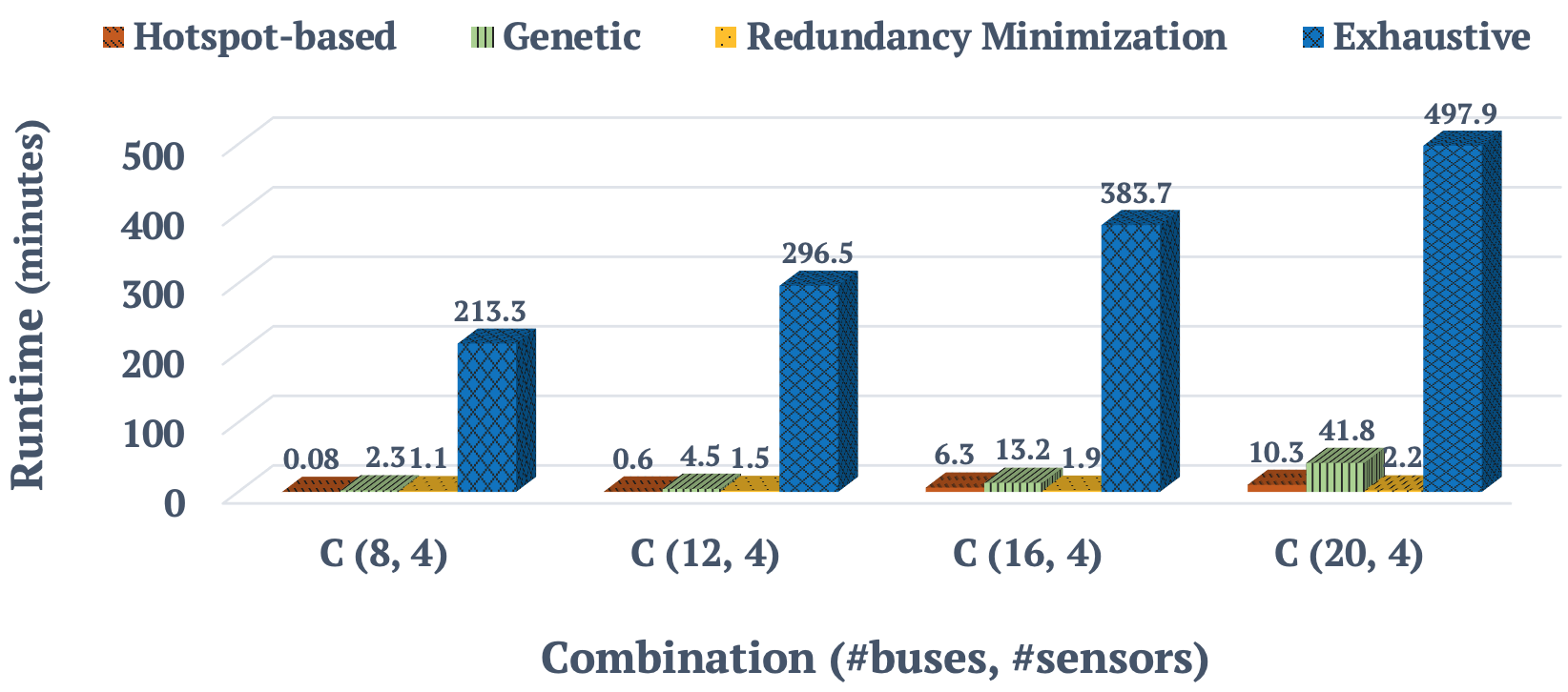} 
    \caption{Runtime Comparison for Different Number of Buses}
    \label{fig:z4runtime}
\end{figure}

Furthermore, we tested the four algorithms on a varying number of buses (i.e., 8, 12, 16, and 20 buses) with four sensors in hand. Figure \ref{fig:z4cvvcompare} shows the CCV earned by each algorithm, and Figure \ref{fig:z4runtime} depicts the runtime comparison of the four algorithms. Likewise, the redundancy minimization algorithm substantially outperforms all other algorithms in terms of the computational cost, while it makes near-optimal selections. We can also observe how the genetic algorithm defeats the hotspot-based algorithm, in terms of the coverage value, by exploring non-hotspot areas and their corresponding buses.

\begin{figure}[htbp]
    \centering
    \includegraphics[width=1\columnwidth]{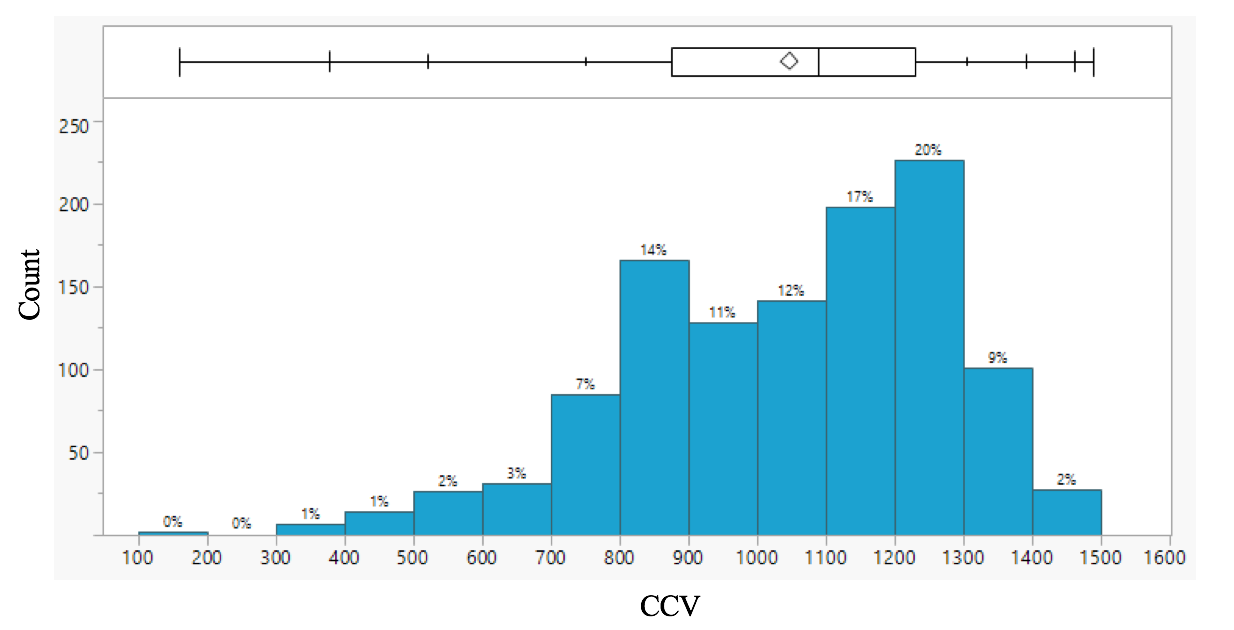} 
    \caption{Results from the Exhaustive Approach}
    \label{fig:exhaustiveApp}
\end{figure}

Now, we provide some sample runs of the algorithms to elaborate on the details of each approach. For instance, Figure \ref{fig:exhaustiveApp} shows the result of our exhaustive algorithm on all the possible bus combinations of 20 buses taken 3 at a time, which is equal to 1140 various combinations. In this figure, the x-axis represents the CCV range, and the y-axis represents the number of bus combinations that belong to each CCV range. We can observe a distribution that is close to the normal distribution in which only two percent of bus combinations earned the highest coverage values.

\begin{figure}[htbp]
    \centering
    \includegraphics[width=0.9\columnwidth]{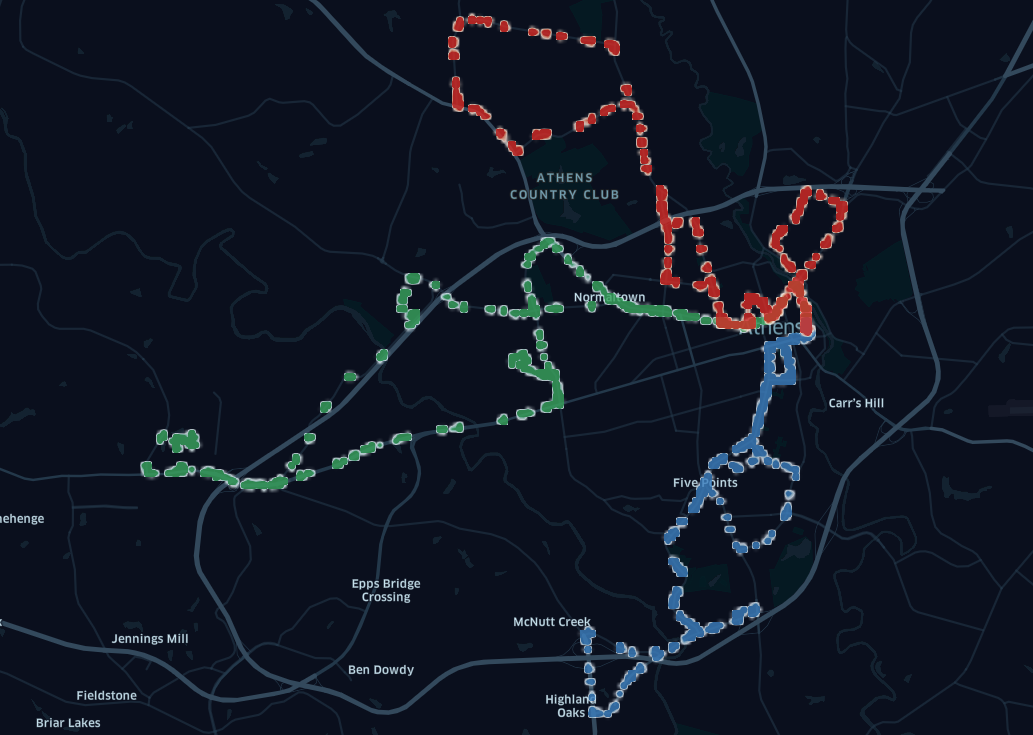} 
    \caption{Trajectory Map of the Bus Combination of the Exhaustive Algorithm }
    \label{fig:Trajectory}
\end{figure}

In this sample experiment, the exhaustive algorithm chooses the most optimal bus combination that gained the highest CCV of 1489. Figure \ref{fig:Trajectory} shows the trajectory data of this bus set on the map, which illustrates how well our proposed objective function was able to select buses with the highest spatiotemporal coverage while having the lowest amount of overlaps. 

\begin{figure}[htbp]
    \centering
    \includegraphics[width=1\columnwidth]{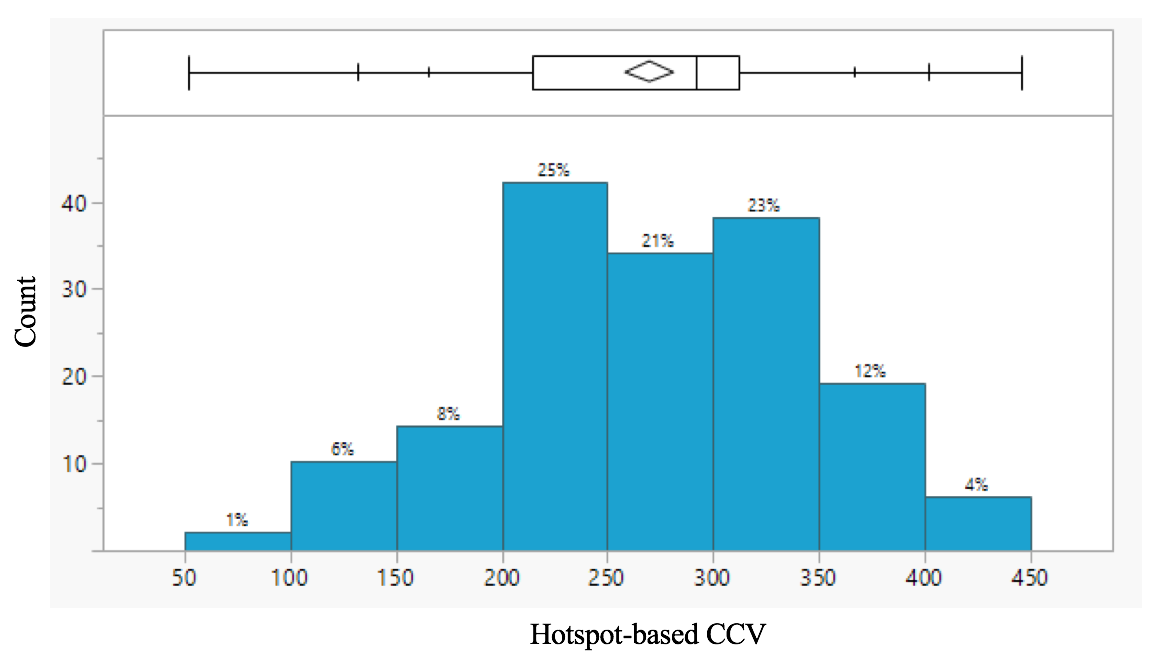} 
    \caption{Results from the Hotspot-based Approach}
    \label{fig:greedy}
\end{figure}

To elaborate on the details of our hotspot-based algorithm, Figure \ref{fig:greedy} depicts the result with the same experimental setup. Considering that this algorithm only focuses on the grid cells that correspond to the hotspot areas, and all non-hotspot cells are assigned to have the weight of zero, the graph shows smaller coverage values. Similarly, we can observe a distribution that is close to the normal distribution, while only four percent of the bus combinations gained the highest coverage values.

\begin{figure}[htbp]
    \centering
    \includegraphics[width=1\columnwidth]{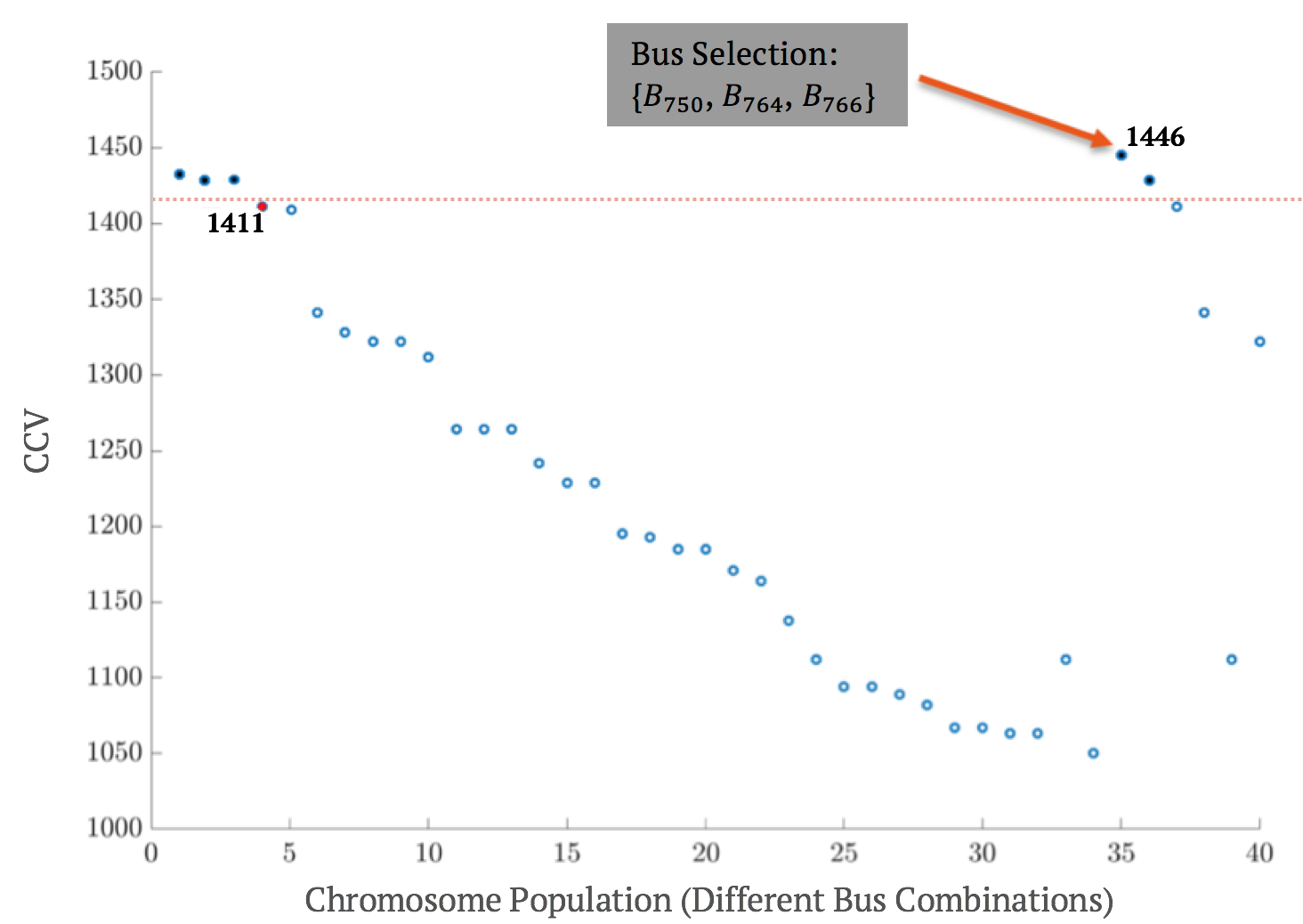} 
    \caption{Visualization of an Example Run of the Utility-Aware Genetic Algorithm}
    \label{fig:genetic}
\end{figure}

Furthermore, Figure \ref{fig:genetic} depicts the coverage values belong to the final population of an example run of our genetic algorithm. In this experiment, we set the population size to be 40. Although we specified the maximum iteration of 20, on average, the algorithm converged after 7 iterations. Our algorithm was able to find 5 bus combinations (black dots) with a higher CCV compared to the best selection in the initial chromosome population (red dot). Therefore, it was able to increase the CCV of 1411, which was earned by the hotspot-based algorithm, to the CCV of 1446. 

Although in this experimental setup, the redundancy minimization algorithm substantially outperforms the genetic algorithm, the spatial resolution, the temporal resolution, and the number of hotspots greatly influence the performance of these two algorithms. For instance, in scenarios where the number of hotspots is minimal, and a high spatiotemporal granularity is expected, the genetic algorithm works much more efficiently. However, our redundancy minimization approach is promisingly the most efficient and effective algorithm to enhance the spatiotemporal coverage in drive-by sensing applications.

\section{Conclusion}
In this paper, we first formulate the problem of choosing an optimal subset of non-dedicated mobile agents on which a limited number of sensors are to be mounted, for the sake of sensing coverage enhancement. Our objective function is implemented by four different algorithms, namely the exhaustive algorithm, the hotspot-based algorithm, the utility-aware genetic algorithm, and the utility-aware redundancy minimization algorithm. Then, we compare their performances and provide experimental results using real trajectory data set of public transportation buses in the city of Athens, GA. We showed how our utility-aware algorithms provide near-optimal solutions and outperform the exhaustive algorithm in terms of runtime. Notably, the redundancy minimization algorithm outperforms other algorithms both in terms of the computation time and the spatiotemporal sensing coverage. Our utility-aware approaches are particularly practical in real-world drive-by sensing platforms where there are some regions with higher relative importance to be consistently monitored.

\bibliography{main.bib}{}
\bibliographystyle{IEEEtran}

\end{document}